\documentclass[usenatbib]{mn2e}
\usepackage{graphicx,natbib}
\usepackage{amsmath,url}

\newcommand{\apj}{ApJ}
\newcommand{\apjl}{ApJ}

\newcommand{\mnras}{MNRAS}

\newcommand{\aap}{A\&A}

\newcommand{\sgra}{Sgr~A$^\star$}

\newcommand{\msun}{{\rm M}_{\odot}}\newcommand{\tmsun}{\mbox{$\msun$}}

\newcommand\simless{\mathbin{\lower 3pt\hbox
   {$\rlap{\raise 5pt\hbox{$\char'074$}}\mathchar"7218$}}}
\newcommand\simgreat{\mathbin{\lower 3pt\hbox
   {$\rlap{\raise 5pt\hbox{$\char'076$}}\mathchar"7218$}}}

\newcommand{\model}[1]{\texttt{#1}}

\title[Cusp influence on the dynamics of stellar discs]
      {Influence of a stellar cusp on the dynamics of young stellar discs and the origin of the S-stars in the Galactic Centre}
      
\author[U. L\"ockmann, H. Baumgardt, \& P. Kroupa]
{
  U. L\"ockmann\thanks{E-mail: uloeck@astro.uni-bonn.de (UL); holger@astro.uni-bonn.de (HB); pavel@astro.uni-bonn.de (PK)}
  , H. Baumgardt\footnotemark[1], and P. Kroupa\footnotemark[1]\\ 
  Argelander Institute for Astronomy, University of Bonn, Auf dem H\"ugel 71, 53121 Bonn,
  Germany\\
}

\begin{document}

\date{Accepted 2009 May 29.  Received 2009 April 27; in original form 2009 February 26}

\pagerange{\pageref{firstpage}--\pageref{lastpage}} \pubyear{2009}

\maketitle

\label{firstpage}

\begin{abstract}
Observations of the Galactic Centre show evidence of one or two disc-like structures of very young stars orbiting the central super-massive black hole within a distance of a few 0.1 pc.
A number of analyses have been carried out to investigate the dynamical behaviour and consequences of these discs, including disc thickness and eccentricity growth as well as mutual interaction and warping.
However, most of these studies have neglected the influence of the stellar cusp surrounding the black hole, which is believed to be $1-2$ orders of magnitude more massive than the disc(s).

By means of $N$-body integrations using our \textsc{bhint} code,
we study the impact of stellar cusps of different compositions.
We find that although the presence of a cusp does have an important effect on the evolution
of an otherwise isolated flat disc,
its influence on the evolution of disc thickness and warping
is rather mild in a two-disc configuration.
However, we show that the creation of highly eccentric orbits strongly depends on the graininess of the cusp (i.e.\ the mean and maximum stellar masses):
While \citet{chang08} recently found that full cycles of Kozai resonance are prevented by the presence of an analytic cusp, we show that relaxation processes play an important role in such highly dense regions and support short-term resonances.
We thus find that young disc stars on initially circular orbits can achieve high eccentricities by resonant effects also in the presence of a cusp of stellar remnants, yielding a mechanism to create S-stars and hyper-velocity stars.

Furthermore, we discuss the underlying initial mass function (IMF) of the young stellar discs and find no definite evidence for a non-canonical IMF.

\end{abstract}


\section{Introduction}

Observations of the Galactic Centre revealed one or two discs of $\approx 6$\,Myr old stars orbiting the central super-massive black hole (SMBH)
at a distance of $\sim$0.1\,pc \citep{lb03,pau06,lu06,lu09,bmf08}.
Until now, a large number of studies have been carried out to investigate the dynamical behaviour of such stellar discs around SMBHs.

\citet{caa08} have shown that a single cold disc in the absence of a perturbing potential cannot explain the observed large inclinations and eccentricities.
\citet{ndcg06} have provided upper mass limits to the discs' masses by studying the amount of warping of two discs at large inclination.
In \citet{lbk08}
we have shown that a significant number of disc stars eventually achieve highly eccentric orbits due to Kozai interaction, thus providing a mechanism to create the observed S-stars and hyper-velocity stars by disruption of close-passage binaries.

All these simulations have been carried out neglecting the stellar cusp present in the vicinity of the SMBH \citep[e.g.,][]{sea+07}.
\citet{chang08} argued that a smooth background potential hinders the creation of high eccentricities in full Kozai cycles since it causes orbital precession.
On the other hand,
\citet{lb09}
have shown that relaxation processes in a grainy stellar cusp surrounding the SMBH act on time scales comparable with the age of the stellar discs
\citep[see also][]{pgma08}. They demonstrate that due
to warping and relaxation, two highly inclined stellar discs cannot be recognised as flat circular discs after 5\,Myr of interaction.
Significant contributions arise from both two-body relaxation due to stellar encounters customary in star clusters, and \emph{resonant} relaxation effects, an enhanced rate of relaxation of angular momentum in near-Keplerian systems \citep{rt96,ha06b,eka08}.
Furthermore, both disc and cusp stars contribute significantly to the disc evolution.

Considering the apparent importance of the influence of a stellar cusp on the disc evolution, in this paper we challenge the results of earlier studies by including both analytic and $N$-body cusp potentials in corresponding models.
In Section~\ref{sec:gcdci:cuspcomp}, we discuss the structure of the stellar cusp as predicted by theory and derived from observations.
Section~\ref{sec:gcdci:models} describes the numerical method underlying our integrations and our models of the stellar discs.
We present our results with respect to the studies mentioned above in Section~\ref{sec:gcdci:results} and discuss them in Section~\ref{sec:gcdci:discussion}.

\section{Structure of the stellar cusp in the Galactic Centre}\label{sec:gcdci:cuspcomp}

The discs of young stars orbiting the SMBH in the centre of the Milky Way are embedded in a population of old stars and stellar remnants.
Theoretical arguments and $N$-body simulations show that a stellar system around an SMBH
evolves into a cusp with a $\gamma = 1.75$ power-law density distribution
\citep{bw76,bme04a,bme04b,pms04,afs04}.
The density profile of the stellar cusp in the Galactic Centre is estimated from the observed luminosity profile and kinematics by \citet{sea+07} as
\begin{equation}
\rho_m(r)=\left(2.8 \pm 1.3\right)\times 10^6\,\msun {\rm pc}^{-3} \left(\frac{r}{0.22\,{\rm pc}}\right)^{-\gamma},
\end{equation}
where $r$ is the distance from the SMBH, $\gamma=1.2$ inside 0.22\,pc and $\gamma=1.75$ outside 0.22\,pc.\footnote{These results are consistent with those of a more recent analysis of \citet{sme09}.}
Hence, the stellar cusp has a mass within 1.5\,pc comparable to the black hole mass.
We use this result (omitting the uncertainty specified) when modelling the cusp as an analytic potential.
However, as we will discuss with the results in Section~\ref{sec:gcdci:results}, relaxation between stars can play an important role in the highly dense central parsec of our Galaxy. Hence it is not sufficient to include such a smooth potential: To perform realistic simulations of disc dynamics in the Galactic Centre, it is required to consider stars and stellar remnants in the $N$-body calculations.

A model of the stellar population in the central 100\,pc of our Galaxy is given by \citet{a05}. Due to dynamical friction causing massive stars to sink towards the centre, it can be assumed that the central few 0.1\,pc, where we also find the young stellar discs, are mass-dominated by the most massive components, namely stellar-mass black holes \citep{fak06,ah08}.

In this paper we will analyse the dynamics of the stellar discs, which will be described in more detail in the next section.
This restricts our interest to the potential within 1\,pc from the SMBH. 
At 2\,pc from the centre \citet{ggw+87} found a rotating ring of gas of $10^6\,\msun$, the so-called circum-nuclear disc (CND).
\citet{ssk08} have shown that the torque exerted by the CND may cause (additional) warping of a stellar disc at $\sim 0.1$\,pc from the SMBH.
Furthermore, it may cause somewhat higher eccentricities due to Kozai interaction (cf.\ Section~\ref{sec:gcdci:sstars}).
Throughout this paper, we will ignore the effect of the CND.
This leaves the stellar cusp as the only dynamical perturbation to the disc evolution.
An exhaustive study of the dynamics of stars in the Galactic Centre will have to include the effects of the CND. However, the general conclusions about the influence of a stellar cusp on the disc dynamics as discussed in this Paper will remain unaffected.

\section{Numerical method and models}\label{sec:gcdci:models}

To test the mutual interaction of two stellar discs orbiting the central SMBH, we performed direct $N$-body integrations using our \textsc{bhint} code\footnote{\textsc{bhint} is freely available
at \url{http://www.astro.uni-bonn.de/english/downloads.php}.}
\citep{lb08}.
\textsc{bhint} has been developed specifically to calculate the dynamics of stars orbiting around an SMBH,
avoiding the large secular error other $N$-body codes based on the Hermite scheme \citep{ma92} accumulate in case of the highly eccentric orbits and extreme mass ratios present in the Galactic Centre.
\textsc{bhint} integrates the equation of motion of a star orbiting an SMBH by moving it along the Keplerian orbit and adding the forces from other components (disc stars and cusp black holes) as perturbations.
It solves Kepler's equation analytically (and thus without secular error) to determine the orbital motion around the dominating SMBH, which is assumed to rest at the origin of the coordinate system, while using the Hermite scheme to integrate the perturbing forces.

\textsc{bhint} includes special treatment of close encounters and fast approaches to allow for high precision.
In addition, it comprises
post-Newtonian (PN) approximation of the effects of general relativity up to order 2.5 following \citet{bla06}, which is added to the perturbations.
However, since the PN terms need relatively high accuracy and thus a significant amount of computational time, we usually switch on these terms only in integrations of highly eccentric orbits, where pericentre-shift becomes important (see Table~\ref{tab:params}). For example,
\citet{lb09} 
have shown that in the regime of the stellar discs, pericentre shift due to the spherical cusp is at least one order of magnitude more effective than the relativistic one, as long as the eccentricity is below $e \approx 0.9$.

\begin{table*}

  \begin{center}
  \begin{minipage}{\textwidth}
    \centering
  \begin{tabular}
  {|cccccccccccccc|}
    \hline
      &&&&&&&&&&&& analytic & \\
      \raisebox{1.5ex}[1.5ex]{Model} & \raisebox{1.5ex}[1.5ex]{$M_1$} & \raisebox{1.5ex}[1.5ex]{$r_1$} & \raisebox{1.5ex}[1.5ex]{$M_2$} & \raisebox{1.5ex}[1.5ex]{$r_2$} & \raisebox{1.5ex}[1.5ex]{$i$} & \raisebox{1.5ex}[1.5ex]{$m$} & \raisebox{1.5ex}[1.5ex]{$\alpha$} & \raisebox{1.5ex}[1.5ex]{$f_{\rm bin}$} & \raisebox{1.5ex}[1.5ex]{$T_{\rm end}$} & \raisebox{1.5ex}[1.5ex]{$M_{\rm SBH}$} & \raisebox{1.5ex}[1.5ex]{$r_{\rm SBH}$} & cusp & \raisebox{1.5ex}[1.5ex]{PN}\\
      \hline
      \model{1a} & 15\,000 & 0.05--0.5 & -- & -- & -- & 1--120 & 1.35 & 30\,\% & 5\,Myr & 15 & 0.22 & -- & -- \\
      \model{1b} & 15\,000 & 0.05--0.5 & -- & -- & -- & 1--120 & 1.35 & 30\,\% & 5\,Myr & -- & -- & yes & -- \\
      \model{1c} & 15\,000 & 0.05--0.5 & -- & -- & -- & 1--120 & 1.35 & 30\,\% & 5\,Myr & -- & -- & -- & -- \\
      \hline
      
      \model{2a} & 10\,000 & 0.05--0.5 & 5\,000 & 0.07--0.5 & 88$^{\circ}$ & 1--120 & 1.35 & 30\,\% & 5\,Myr & 15 & 0.22 & -- & -- \\
      \model{2b} & 10\,000 & 0.05--0.5 & 5\,000 & 0.07--0.5 & 88$^{\circ}$ & 1--120 & 1.35 & 30\,\% & 5\,Myr & -- & -- & yes & -- \\
      \model{2c} & 10\,000 & 0.05--0.5 & 5\,000 & 0.07--0.5 & 88$^{\circ}$ & 1--120 & 1.35 & 30\,\% & 5\,Myr & -- & -- & -- & -- \\
      \hline
      
      \model{3a} & 18\,000 & 0.05--0.5 & 9\,000 & 0.07--0.5 & 88$^{\circ}$ & 1--120 & 2.3 & 40\,\% & 5\,Myr & 15 & 0.22 & -- & -- \\
      \model{3b} & 13\,000 & 0.05--0.5 & 8\,500 & 0.07--0.5 & 105$^{\circ}$ & 1--120 & 2.3 & 13\,\% & 5\,Myr & 15 & 0.22 & -- & -- \\
      \model{3c} &  11\,000 &  0.05--0.5 &  6\,500 &  0.07--0.5 &  88$^{\circ}$ &  1--120 & 2.3 &  13\,\% &  4\,Myr &  15 &  0.22 &  -- &  -- \\
      \model{3d} &  24\,000 &  0.05--0.5 &  12\,000 &  0.07--0.5 &  88$^{\circ}$ &  0.01--120 & canonical &  30\,\% &  4\,Myr &  -- &  -- &  -- &  -- \\
      \hline
      
      \model{4a} & 20\,000 & 0.05--0.5 & 10\,000 & 0.07--0.5 & 125$^{\circ}$ & 1--120 & 1.35 & 30\,\% & 6\,Myr & 15 & 0.22 & -- & if $e>0.9$ \\
      \model{4b} & 15\,000 & 0.05--0.5 & 7\,500 & 0.07--0.5 & 88$^{\circ}$ & 1--120 & 1.35 & 40\,\% & 6\,Myr & 15 & 0.22 & -- & -- \\
      \model{4c} & 20\,000 & 0.03--0.4 & 8\,000 & 0.05--0.4 & 100$^{\circ}$ & 1--120 & 2.3 & 40\,\% & 6\,Myr & 15 & 0.22 & -- & -- \\
      \model{4d} & 15\,000 & 0.05--0.5 & 7\,500 & 0.07--0.5 & 88$^{\circ}$ & 1--120 & 1.35 & 40\,\% & 6\,Myr & 40 & 0.22 & -- & yes \\
      \hline
    \end{tabular}
    \caption
    {Initial parameters used in our models. These columns give the following in order:
    Model name; mass of the first disc; radial extent of the first disc; mass of the second disc; radial extent of the second disc;
    relative inclination of the discs; mass range and IMF slope of disc stars; binary fraction; simulated time; mass of the cusp SBHs; radial extent of SBH cusp; analytic cusp potential; post-Newtonian treatment for disc stars.
    All masses are in solar masses, all distances in parsec.
    Where present, both SBH   and analytic cusps follow the density profile derived by \citet{sea+07}.
    For model \model{3d} we assumed the multi-part power-law canonical IMF down to 0.01\,\tmsun;
    to keep the total number of stars low, we had to replace the low-mass stars ($<1\,\msun$) by 1\,\tmsun\ stars of equal total mass.
    The difference between the canonical and $\alpha=2.3$ models is that the latter are restricted to $1-120\,\msun$.
      \label{tab:params}}
    \end{minipage}
  \end{center}
\end{table*}

\begin{figure*}
  \begin{center}
    \includegraphics[width=\textwidth]{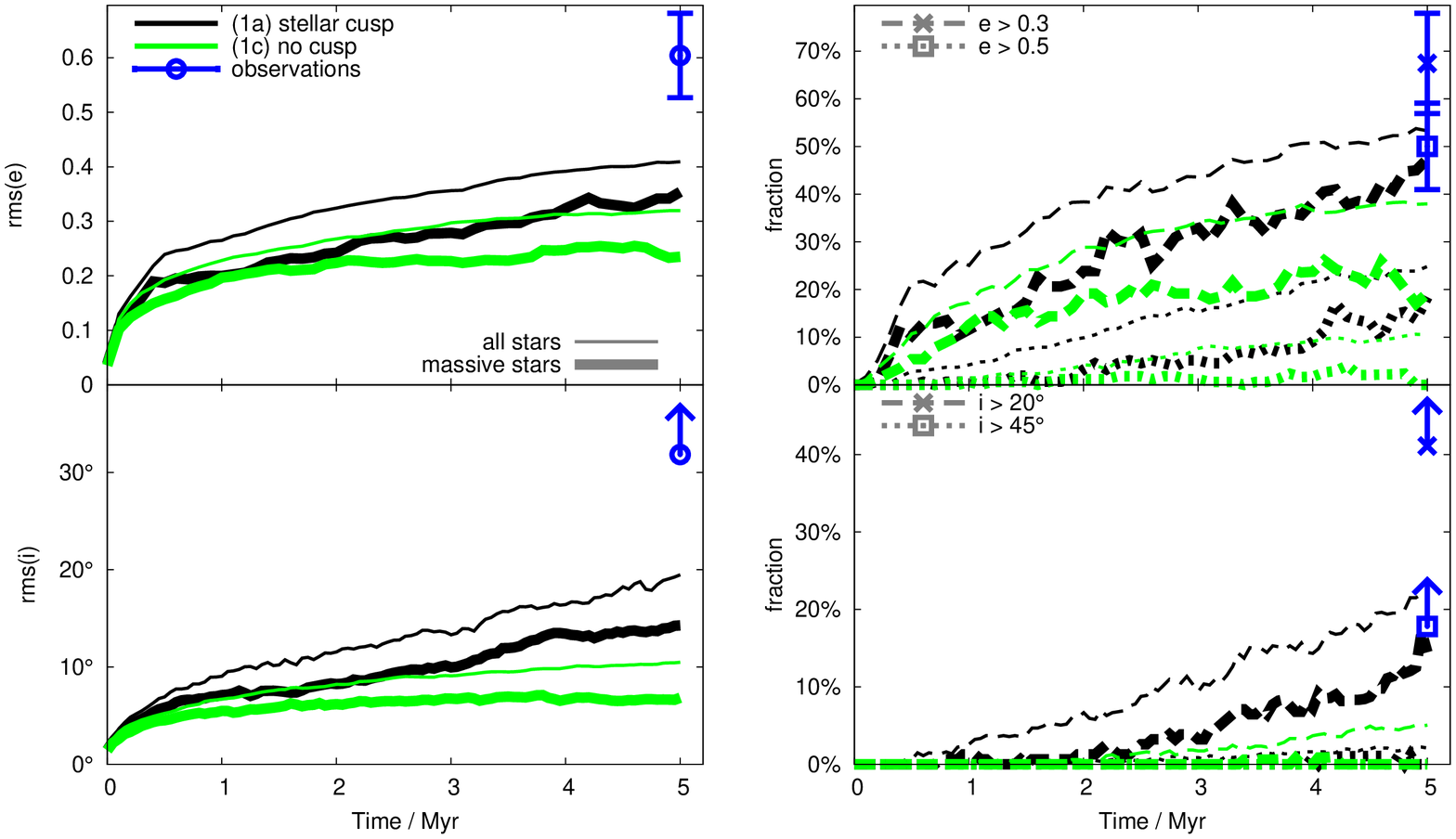}
  \end{center}
  \caption[Time evolution of orbital eccentricity and inclination of a cold disc]{Time evolution of orbital eccentricity (top panels) and inclination (bottom panels) of an initially cold disc.
    The left panels show the root mean square values for models in the presence (\model{1a}; black) and absence (\model{1c}; green) of a stellar cusp.
    The corresponding dashed and dotted lines in the right panels show the fraction of stars on orbits with eccentricities larger than 0.3 and 0.5 (top),
    and inclinations larger than 20$^{\circ}$ and 45$^{\circ}$ relative to the fitted disc plane (bottom), respectively.
    While the thin lines refer to the distribution of all stars, the thick lines represent the massive stars ($>20\,\msun$) only.
    In line with the results of \citet{pgma08}, we find a faster increase of eccentricities and inclinations in the presence of a stellar cusp.
    For comparison, the blue points indicate the corresponding values for the young disc stars observed in the Galactic Centre \citep{pau06}, which have masses $>20\,\msun$.
    We assign Poisson error bars to the eccentricity values.
    Since no estimates are provided for the orbital planes, we use disc fitting of the velocity vectors for a very pessimistic lower limit of the inclination values.
    \label{fig:1disc-e-i}}
\end{figure*}

Unless stated otherwise, our model parameters are as follows:
We start with two cold stellar discs around a $3.5\times 10^6\,\msun$ SMBH at high relative inclination $i$.
The discs have finite radial extents $r_{1,2}$, a surface density $\Sigma (r) \propto r^{-2.5}$, and a thickness (standard deviation) of 1.44$^{\circ}$. The stellar orbits have a mean eccentricity of 0.03.
We vary the initial values for disc masses, inclination, and radial extent within ranges compatible with the observations (see Table~\ref{tab:params}).

The underlying IMF in most of our models has a slope of $\alpha = 1.35$,
which was derived by \citet{pau06} for the stellar discs in the Galactic Centre.
This is considerably flatter than the canonical initial mass function (IMF) according to \citet{k01},
$\xi\left(m\right)\propto m^{-\alpha_i}$, with $\alpha_0=0.3$ $\left(0.01\le m/\msun<0.08\right)$, $\alpha_1=1.3$ $\left(0.08\le m/\msun<0.5\right)$, and $\alpha_2=2.3$ $\left(0.5\le m/\msun\right)$.
The number of stars in the mass interval $m$ to $m+{\rm d}m$ is $\xi(m) {\rm d}m$, and the \citet{s55} IMF has $\alpha=2.35$.
Since the disc mass for an IMF as flat as $\alpha = 1.35$ is not very sensitive to the lower stellar mass limit, we use a mass range of $1-120\,\msun$.

For most of our models, we assume a disc age of 5\,Myr \citep{pau06}.
The discs cannot be much older: According to standard stellar evolution models \citep[e.g.][]{hpt00}, an age of 6\,Myr implies a maximum zero-age main sequence (ZAMS) mass of 32\,\tmsun\ for observed stars. The observed number of disc stars with minimum ZAMS masses of 20\,\tmsun\ \citep{pau06} would thus imply a total initial disc mass of 16\,750\,\tmsun\ for the IMF with $\alpha=1.35$ (77 stars with $20 \le m/\msun \le 32$), plus some binary fraction, and the discs are believed to have a total mass $\simless 15\,000\,\msun$ \citep{ndcg06}. However, we vary the disc masses and ages consistently in Sections~\ref{sec:gcdci:imf} and \ref{sec:gcdci:sstars}.

To test for disruption of binaries, we mark a certain fraction $f_{\rm bin}$ of stars in our models as binaries (varying between the different models, see Table~\ref{tab:params}), and double their mass (assuming equal-mass binaries). We track close encounters during our simulations to determine possible points of disruption.
Disruption of a binary of mass $m_{\rm bin}$ and separation $a$ occurs if it passes another particle (SMBH, disc or cusp star) of mass $M$ at a distance less than $R_{\mathrm{tid}} \sim a \sqrt[3]{M/m_{\rm bin}} \label{eq:r-tid-bin}$ \citep{yt03}.

We accounted for the influence of a background cusp as discussed in Section \ref{sec:gcdci:cuspcomp} either by using an analytic potential, or by adding 14\,000 stellar black holes (SBHs) of 15\,\tmsun\ into the inner 0.22\,pc, consistent with the density profile derived by \citet{sea+07}.

Table~\ref{tab:params} lists the varying parameters of the models used throughout this paper.
On 3\,GHz desktop computers equipped with GRAPE6-Pro/8 or MicroGRAPE boards, calculation of these models took from five hours (top-heavy IMF, no SBH cusp) to a few 100 hours (all other models) per model.

\section{Results}\label{sec:gcdci:results}

\subsection{Evolution of a single disc}
\citet{caa08} have shown that a single cold disc in the absence of a perturbing potential cannot explain the observed large inclinations and eccentricities.
The required perturbing potential may result from a massive circum-nuclear disc (\citealp{ght94}; cf.\ \citealp{ssk08}), an intermediate-mass black hole and/or star cluster \citep[probably IRS 13;][]{mpsr04}, or a second disc (\citealp{pau06}; cf.\ 
\citealp{lb09}%
).

To test whether a spherical cusp alone significantly alters the evolution of a disc, we integrated models of a single disc with the $\alpha = 1.35$ IMF suggested by \citet{pau06}, restricted to $1-120\,\msun$.\footnote{Using a Salpeter/Kroupa IMF \citep{k01} with $\alpha = 2.3$ for respective masses lead to similar results. For a discussion on the choice of underlying IMF see Section~\ref{sec:gcdci:imf}.}

Figure~\ref{fig:1disc-e-i} shows the time evolution of eccentricity and inclination of a single disc in a stellar cusp (\model{1a}).
In line with the results of \citet{pgma08}, we find a faster increase of eccentricities and inclinations than in the absence of a stellar cusp (\model{1c}). This still holds if we consider only the massive stars. However, less than 20 per cent of the massive stars (above $20\,\msun$) have final eccentricities $e>0.5$, and a mere two per cent of the massive stars reach an inclination larger than 45$^{\circ}$, while half of the disc stars observed have eccentricities $e>0.5$, and a large number of stars have much higher inclinations to the observed clockwise disc \citep{pau06}.
In fact, the largest contribution to the high values of observed eccentricities and inclinations stems from the counter-clockwise stars.

We conclude that the stellar cusp does have a significant effect on the evolution of eccentricities and especially orbital inclinations,
but agree with \citet{caa08} in that it cannot account for the highly eccentric and counter-clockwise stars assuming an initial configuration of a single cold disc in the absence of any perturbing potential as discussed above.

\subsection{Evolution of two interacting discs}
\citet{ndcg06} have shown how the mutual interaction of two discs with large relative inclination leads to an increase of disc thickness.
The evolution of orbital eccentricities and inclination (i.e.\ disc warping) in the presence of a cusp has been studied in detail
by \citet{lb09}. They find
that while the evolution of orbital inclinations is dominated by the discs' potential (precession and warping), the change in eccentricity and semi-major axis of the disc stars is driven by non-resonant as well as resonant relaxation with disc and cusp stars.

Here we analyse the effect of the stellar cusp on the discs by computing the discs modelled by \citet{lb09} a) in a cusp of 15\,\tmsun\ SBHs, b) in an analytic cusp, and c) in the absence of any cusp potential.

\begin{figure}
  \begin{center}
    \includegraphics[width=8.3cm]{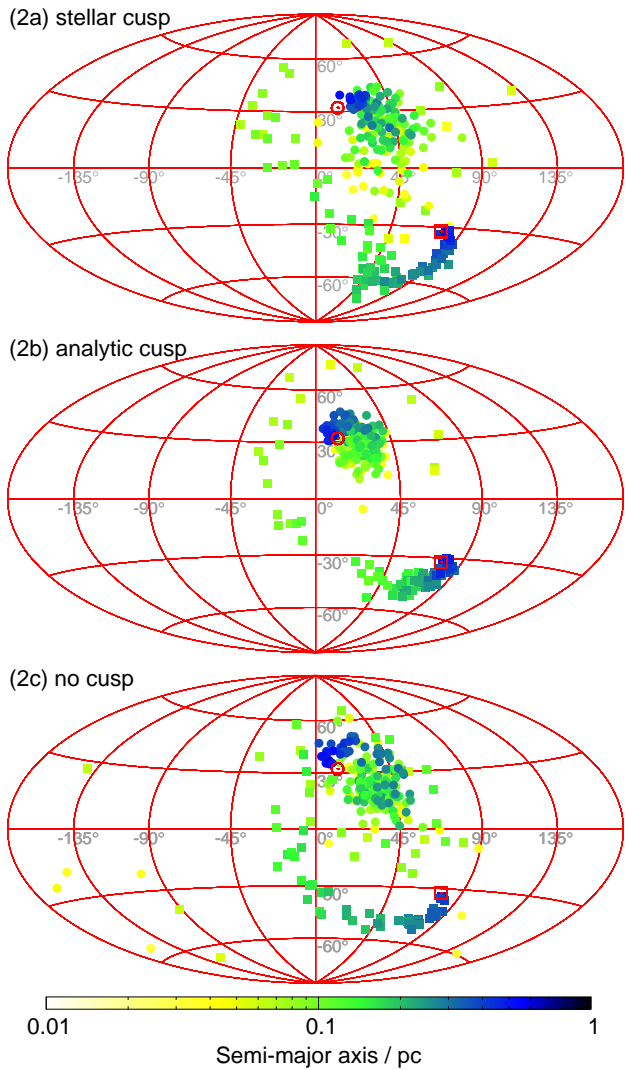}
 \end{center}
  \caption[Cusp influence on the orientation of stellar orbits]{Aitoff map of the orientation of orbital planes after an integration time of 5\,Myr. The panels show the results of a model including stellar cusp, analytic cusp, and no cusp (top to bottom). Filled circles and squares denote the normalised angular momentum vectors of stars from the clockwise and counter-clockwise disc, respectively. The symbols are colour-coded by semi-major axis $a$; the open (red) symbols show the mean angular momentum vector of the initial discs. The direction and amount of warping does not depend significantly on the presence of a cusp.
    \label{fig:warp}}
\end{figure}

\begin{figure*}
  \begin{center}
    \includegraphics[width=\textwidth]{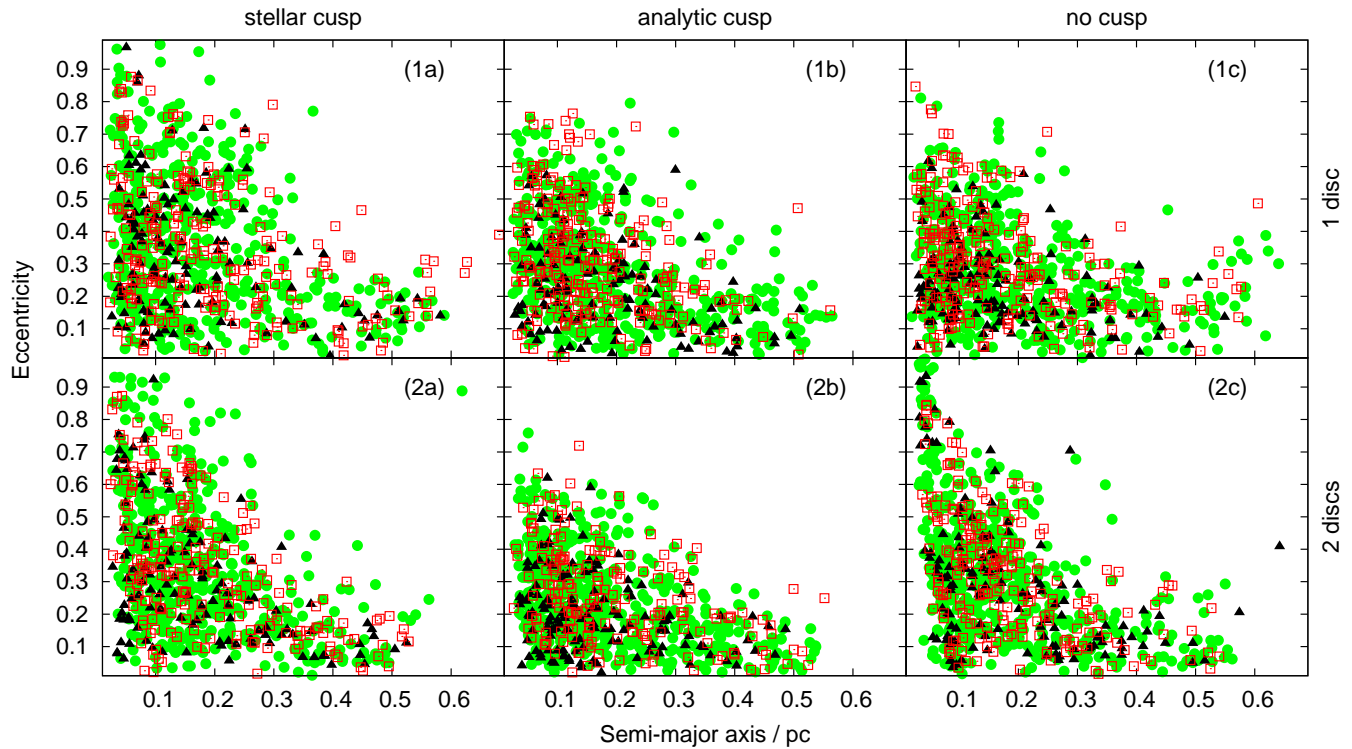}
 \end{center}
  \caption[Cusp influence on eccentricity distribution]{Eccentricity distribution as a function of semi-major axis after an integration time of 5\,Myr.
  The panels show the results of models of one disc (top) and two discs (bottom) including stellar cusp, analytic cusp, and no cusp (left to right).
  The stars are grouped by final mass above 15\,\tmsun\ (black triangles), below 2\,\tmsun\ (open red squares), and in between (green circles).
  High eccentricities are only achieved in the presence of a stellar cusp, or by interaction between the two discs in the absence of a spherical potential.
    \label{fig:eacomp}}
\end{figure*}

Figure~\ref{fig:warp} shows the orientation of the orbital planes of the disc stars after an integration time of 5\,Myr. 
Since precession is dominated by the disc potentials, direction and amount of warping do not depend significantly on the cusp.
On the other hand, the distribution of eccentricities does depend on the surrounding cusp:
Figure~\ref{fig:eacomp} shows the final distribution of the disc stars' eccentricities as a function of semi-major axis, for both single-disc and two-disc setup.
While the overall eccentricity distribution is similar in all cases, eccentricities $e>0.8$ are only achieved in the presence of a stellar cusp, or in a two disc setup without any cusp potential (at the very inner edge). As we will discuss in Section~\ref{sec:gcdci:sstars}, while high eccentricities reached by long-term Kozai interaction with the two discs
are impeded by an analytic cusp, relaxation processes in a stellar cusp
\citep[cf.][]{lb09}
can lead to eccentricities high enough for short-term resonances to be effective despite the spherical potential, thus explaining the high eccentricities in three of our six cases (two discs in the absence of a cusp, or one or two discs interacting with cusp remnants).

\subsection{Constraining the disc mass and IMF}\label{sec:gcdci:imf}
As a direct result of the disc warping described in the previous section, \citet{ndcg06} have provided upper mass limits to the discs' masses considering their apparent flatness as measured by \citet{pau06} after a few Myr of interaction. However, \citet{lu09} did not find convincing evidence for a counter-clockwise disc as flat as 19$^{\circ}$, which can also be interpreted as an indication for a higher mass of the massive disc in a two-disc model.
\citet{lb09}
have extended the argument of warping as a mere measure of disc thickening by measuring the warping angle as a function of radial distance and found good agreement of a two-disc model with the observations, showing how the warping angle can reach 50$^{\circ}$ already at 0.15\,pc from the SMBH.

\begin{figure}
  \begin{center}
    \includegraphics[width=8.3cm]{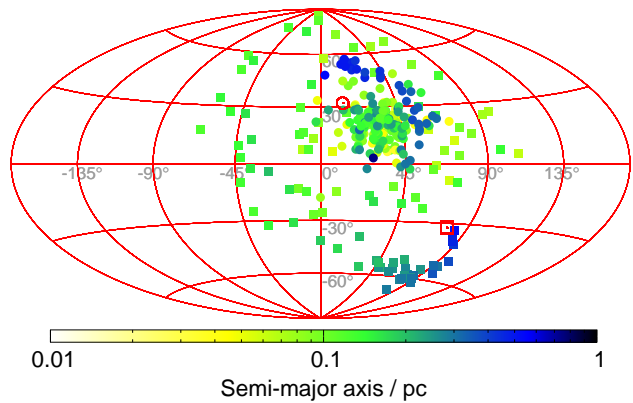}
 \end{center}
  \caption[Final orientation of simulated stellar orbits of massive discs]{Orientation of orbital planes of model \model{3a} following a Kroupa IMF above 1\,\tmsun\ after an integration time of 5\,Myr. Despite significantly higher disc masses, the overall trend of disc warping is comparable to the results displayed in Fig.~\ref{fig:warp}.
    \label{fig:imf1}}
\end{figure}

Here we calculated different models with two discs of stars following a canonical IMF from $1-120\,\msun$. Using binary fractions of 40 and 13 per cent implies total masses of 27\,000\,\tmsun\ (\model{3a}) and 21\,500\,\tmsun\ (\model{3b}), respectively, to have the same number of massive stars.
Figure~\ref{fig:imf1} shows the orientation of orbital planes of model \model{3a} after 5\,Myr of integration, which shows the same trend as the results of the previous section; the same holds for model \model{3b}.
Assuming a disc age of 4\,Myr \citep[the lower limit as estimated by][]{pau06} reduces the initial disc mass by 20 per cent (\model{3c}), since a larger fraction of massive stars survives the shorter disc age, and thus a smaller initial number of massive stars is needed to explain the observed 77 stars. In another model (\model{3d}), we choose the initial disc mass to be consistent with the canonical IMF \citep{k01} down to 0.01\,\tmsun, resulting in 80 per cent more mass than with a cut-off at 1\,\tmsun. To keep the total number of stars low, we had to replace the low-mass stars ($<1\,\msun$) by 1\,\tmsun\ stars of equal total mass.
Again, the overall spiral trend of disc warping as plotted in Fig.~\ref{fig:imf1} is similar to the earlier results.

\begin{figure}
  \begin{center}
    \includegraphics[width=8.3cm]{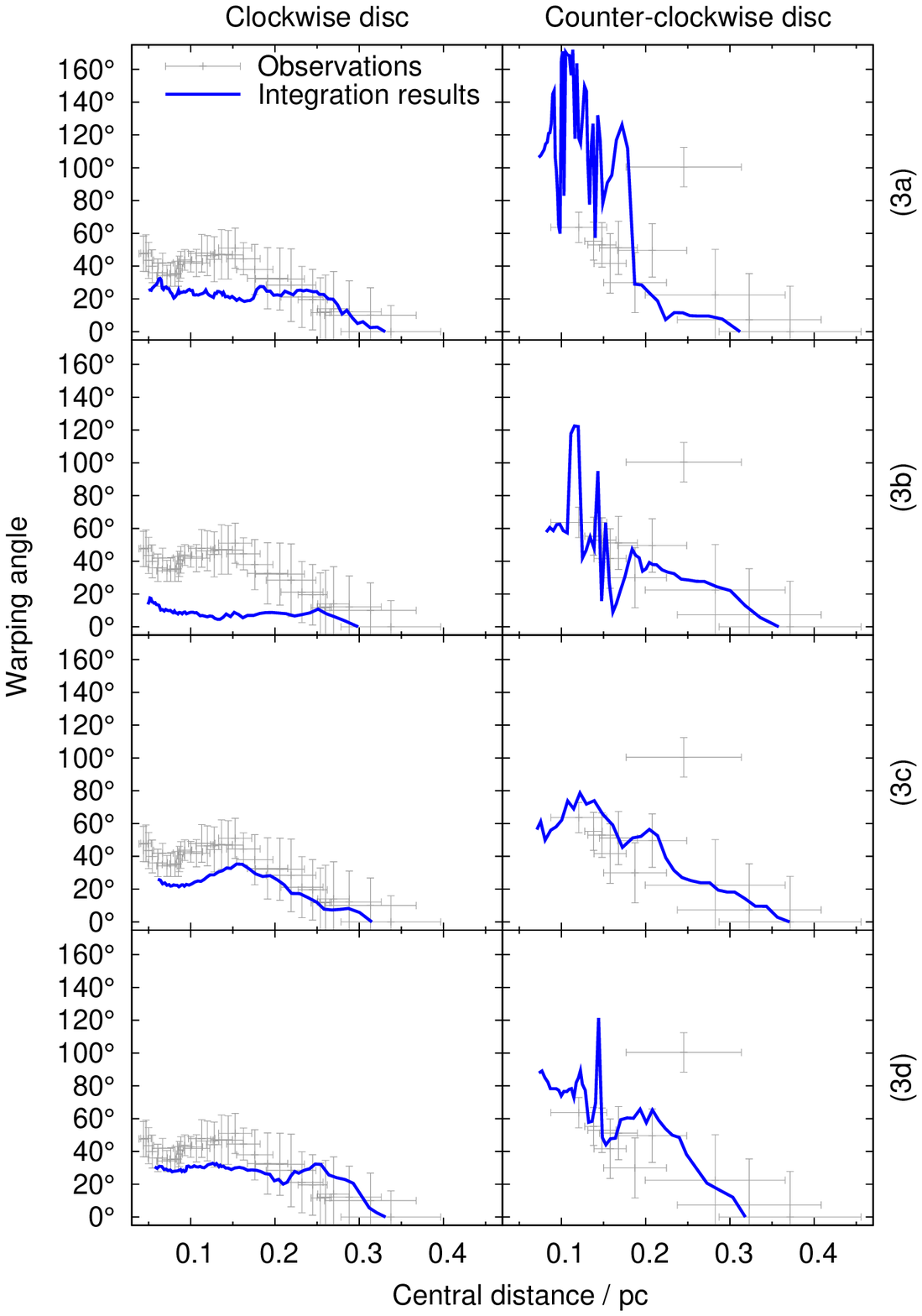}
 \end{center}
  \caption[Disc warping angle as a function of central distance]{Disc warping angle as a function of central distance. We plot the angle between the plane fitted to any disc's subset of stars and the outermost fitted plane for the respective disc. The left panels show the results for the clockwise disc from observations (grey error bars) as well as for models \model{3a} through \model{3d} (blue lines, top to bottom);
  the right panels show the respective results for the counter-clockwise systems.
  Horizontal and vertical error-bars indicate standard deviation of central distance and 1$\sigma$-thickness of the fitted planes, respectively.
  The corresponding error bars for the simulations are generally smaller and thus omitted for clarity
  \citep[cf.][]{lb09}.
  In none of our models does the amount of warping of the clockwise disc exceed the observed value.
  The modelled counter-clockwise discs dissolve towards the inner part, where in most cases warping is no longer a smooth function of central distance.
    \label{fig:imf2}}
\end{figure}

Figure \ref{fig:imf2} shows the angle between a disc fitted to massive stars at a certain radius and the outermost fitted disc.
While the amount of warping of all clockwise (massive) discs modelled with a canonical IMF is compatible with the observed values
\citep[or even below; cf. discussion in][]{lb09},
the integration results of the counter-clockwise disc fit the observations well only outside $\approx 0.16$\,pc from the SMBH.
Inside this distance, the counter-clockwise disc dissolves, and thus the warping angle is no longer a smooth function of central distance.
However, it is essential to note that the observed stars are assigned to one of the two discs depending on whether they are moving clockwise or counter-clockwise in the sky, since there is no other way to tell their disc of origin. This implies that none of the squares on the northern hemisphere of Fig.~\ref{fig:imf1} would be considered for determining the warping angles of the counter-clockwise disc. Furthermore, more stars observed within 0.16\,pc from \sgra\ move on projected radial orbits (and are thus completely dismissed) than on projected counter-clockwise orbits. If these (and even some of the clockwise-moving stars) stem from the more strongly warped counter-clockwise disc, they might significantly increase the amount of warping of the observed counter-clockwise disc.
We waive a further analysis since assignment of stars to one of the two discs is rather uncertain in the inner 0.16\,pc with the currently available observational data. In addition, the choice of the line of sight in our models is arbitrary.

Considering this and the fact that there is as yet no model explaining all of the observed features of the young stars and including all perturbations like e.g.\ the CND,
we conclude that these models are still consistent with the observations.
This seems to be in contrast to the results of \citet{ndcg06}, however they used stellar discs only out to 0.2\,pc from the SMBH, while more than one third of the stars listed by \citet{pau06} are further away from \sgra.
In addition, they did not distinguish between stars in the outer parts of the discs which almost retain their orbital planes, and stars close to the centre which strongly precess.

Our results show
that a canonical IMF over the range $0.01-120\,\msun$ cannot be excluded dynamically for the young discs in the Galactic Centre.
\citet{pau06} find a flat IMF from analysing the $K$-band luminosity function of the disc stars; however, they assume a lower mass limit of 20\,\tmsun\ for the observed stars, and their estimate of the disc age of $\approx 6$\,Myr limits the initial masses of the now observed stars to $\approx 32\,\msun$, so this slope refers to a fairly small mass interval only.
\citet{mmt07} find a best fit of observations in the central parsec of our Galaxy with a model of continuous star formation with a top-heavy IMF.
Furthermore, \citet{ns05} argue that the X-ray luminosity of the \sgra\ field is one order of magnitude too low to account for the number of young $\simless\, 3\,\msun$ stars expected from a canonical IMF, considering the large number of O-stars observed in the discs. This can also be explained by a higher low-mass cutoff of the IMF at of the order of 1\,\tmsun.
\citet{aba06} find that eccentricities as high as $0.2-0.3$ can only be explained by a top-heavy IMF when assuming a circular disc origin, however they neglect the influence of the cusp and the second disc as elaborated
by \citet{lb09};
both suggest an eccentric disc origin for the counter-clockwise system.
Simulated star formation of gaseous discs suggests that stars form with a top-heavy IMF only if the cooling times are just as large as allowed to still induce disc fragmentation (\citealp{ncs07} and references therein).
Unfortunately, theoretical IMF predictions have failed in the past to correctly describe the observations near the Galactic Centre \citep{k08}.

Another argument for a top-heavy IMF arises from the relation between disc and cusp mass.
Assuming that we do not live in a special time, the formation of stellar discs in the Galactic Centre must be a frequent event.
To explain the observed number of $\ge 20\,\msun$ disc stars at an assumed age of 5\,Myr with the $\alpha = 1.35$ IMF suggested by \citet{pau06}, a total initial disc mass of $\sim 15\,000\,\msun$ is required, of which a mass of $\sim 3000\,\msun$ ends up in stellar remnants.
Using the canonical IMF requires a total initial disc mass of $\sim 45\,000\,\msun$, of which $\sim 30\,000\,\msun$ in remnants and low-mass stars are retained over a Hubble time.
Assuming that
most stars survive in the central region (or are replaced by inspiralling massive remnants), the observed cusp mass of 
$\sim 750\,000\,\msun$ in the central 0.5\,pc \citep{sea+07} allows for the formation of discs like those observed every 50\,Myr for a top-heavy IMF, and only every 500\,Myr for a canonical IMF.
The fact that we observe very young stellar discs thus seems much more unlikely if star formation in the Galactic Centre follows the canonical IMF.
However, it may also be an indication of recently enhanced star formation processes. For example, if the Galactic Bar is young, as an increased fraction of barred galaxies for lower redshifts suggests \citep{see+08}, bar-induced gas inflow may explain such an enhancement (\citealp{sw93} and references therein).

We conclude that while theoretical star formation arguments and some observations indicate a top-heavy IMF in the centre of the Milky Way, our self-consistent models of stellar dynamics allow the existence of a canonical IMF as well. The observational evidence clearly needs to be improved by star counts, while the theory of star formation is currently not able to make conclusive statements on the IMF of stars in the Galactic Centre (see \citealp{k08} for a discussion of recent failures of IMF theory).

\subsection{Creation of S-stars in a cusp environment\label{sec:gcdci:sstars}}
In \citet{lbk08}
we have shown that a significant number of disc stars eventually achieve highly eccentric orbits due to Kozai interaction, thus providing a natural mechanism to create the observed S-stars and hyper-velocity stars by disruption of close-passage binaries.
However, \citet{ips05} remarked that the effects of pericentre shift due to relativistic precession and a spherical cusp can destroy the Kozai mechanism.
While the relativistic precession period decreases strongly with increasing orbital eccentricity,
\begin{equation}
  P_{\rm prec,GR} = 2 \times 10^5 P_{\rm orb} \left(\frac{a}{0.1\,\rm pc}\right)\left(1-e^2\right)
\end{equation}
(where $P_{\rm orb}$, $a$ and $e$ are the orbital period, semi-major axis, and eccentricity, respectively),
they showed that cusp precession at a given distance is most effective for near-circular orbits:
\begin{equation}
  P_{\rm prec,cusp} = 73 P_{\rm orb} \left(\frac{a}{0.1\,\rm pc}\right)^{-1.8}\left(1-e^2\right)^{-0.5}
\end{equation}
(both equations assuming $M_{\rm SMBH}=3.5\times 10^6\msun$ and the cusp profile inside 0.22\,pc by \citealp{sea+07}).
From this we conclude that the very high eccentricities in Kozai resonance are prevented by relativistic precession at high eccentricities and by cusp precession at low eccentricities.

\begin{figure}
  \begin{center}
    \includegraphics[width=8.3cm]{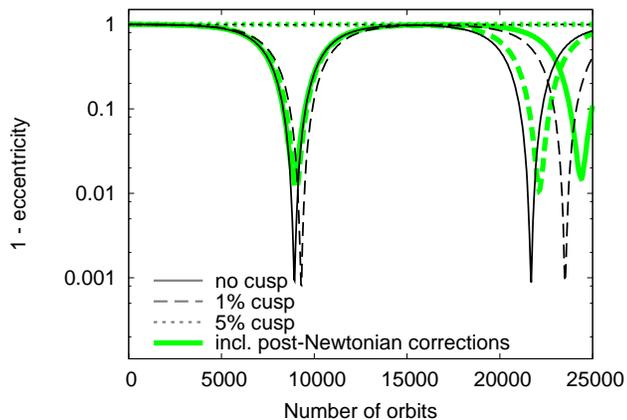}
  \end{center}
  \caption[Influence of stellar cusp and relativistic effects on Kozai resonance]{Eccentricity evolution of a test star undergoing Kozai resonance driven by a fictious 15\,000\,\tmsun\ particle representing the disc potential. Both particles have initially circular orbits about the $3.5\times 10^6\,\msun$ SMBH with semi-major axes of 0.04\,pc and 0.16\,pc, respectively, resulting in the star's orbital period of 400\,yr.
  We calculate models without cusp (solid black line), one per cent (dashed line), and five per cent (dotted line) of the extended cusp mass observed in the Galactic Centre (modelled as a smooth potential).
  Each curve is accompanied by a corresponding thick green curve which represents a respective integration including post-Newtonian terms up to 2.5 PN to account for the effects of general relativity.
  While relativistic effects damp the Kozai effect at high eccentricities, the presence of a stellar cusp leaves it almost unchanged until a mass threshold of a few per cent of the observed value, where eccentricity growth is stopped at very small values so that the orbit remains almost circular (straight dotted lines).
    \label{fig:kozai}}
\end{figure}

Figure~\ref{fig:kozai} shows the eccentricity evolution during Kozai resonance for integrations of a three-body problem with and without a cusp, and with and without PN treatment for relativistic effects. It is clearly seen that the PN terms prevent higher eccentricities when a certain eccentricity threshold is reached (in this case $e\approx 0.9$), while their effect on the rest of the Kozai cycle is not dramatic.\footnote{The minor changes in oscillation period caused both by cusp and PN terms reflect the small perturbations to the Kozai cycle due to precession, where resonance is slightly out of phase with the unperturbed case.}
This can be understood by general relativity being most effective for high eccentricities (and thus small pericentre distances). On the other hand, the changes on both maximum eccentricity and Kozai period exerted by a stellar cusp are quite low, until a certain mass threshold of the cusp is reached, which prevents not only the highest eccentricities but essentially any significant eccentricity increase and thus the complete Kozai cycle. Even for the innermost disc stars ($a\approx 0.03$\,pc), we find this threshold to be at only a few per cent of the cusp mass given by \citet{sea+07}:
While the 1 per cent curve reaches $e=0.99$, the 5 per cent curve does not exceed $e=0.01$.

As a result of this analysis, we cannot expect any disc star in the Galactic Centre to undergo a full Kozai cycle, which was also shown by \citet{chang08}. However, if a star achieves high eccentricity by (non-)resonant relaxation or close encounters
\citep{lb09},
it may find itself in a point of Kozai resonance where the time until maximum eccentricity is significantly shorter than the full Kozai period, and also the cusp precession period is longer than for the circular case.
Furthermore, in our system of two disc potentials and a large reservoir of perturbing disc and cusp stars, other forms of resonance may occur (short-term Kozai/resonant relaxation with changing sources of perturbation).

Going beyond
\citet{lbk08},
we computed different models of two-disc systems in the presence of a stellar cusp. As a result of cusp precession, we find only $1-2$ binaries in each model achieving eccentricities high enough for binary disruption to occur (\model{4a}, \model{4b}).
When using a canonical IMF down to 1\,\tmsun, 7 S-stars are created (\model{4c}). Besides the smaller disc extent used for model \model{4c} (resulting in faster relaxation and shorter Kozai periods), this is due to the number of B-stars being four times higher than in the previous cases (and also the total mass increase), while the number of O-stars is still consistent with the observed value. Due to the increase of total disc mass, the number of S-stars created may increase even further when using a canonical IMF down to 0.01\,\tmsun.

For those stars achieving high eccentricities we can see the behaviour as predicted above: The eccentricity changes in a random fashion by relaxation processes until a certain value of high eccentricity is achieved. At this point, the star (or binary) goes into Kozai resonance for a time scale long enough to achieve extreme eccentricity, but too short to be prevented by precession effects.

\begin{figure}
  \begin{center}
    \includegraphics[width=8.3cm]{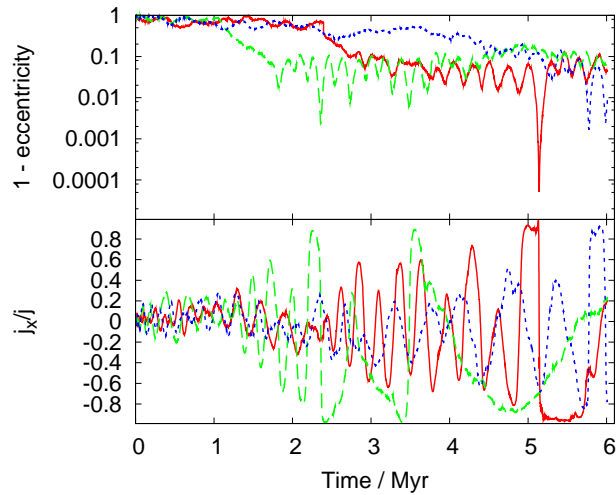}
  \end{center}
  \caption[Evolution of orbital eccentricity and orientation for selected stars]{Evolution of orbital eccentricity (top) and $x$ component of the normalised angular momentum vector, $j_x/j$ (bottom), of three selected stars of model \model{4b}.
  The cyclic nature of eccentricity evolution as well as the rapid change of the stars' orbital planes as indicated by $j_x/j$ are revealing of the Kozai mechanism.
    \label{fig:eccevol}}
\end{figure}

Figure~\ref{fig:eccevol} shows the evolution of eccentricity as well as one component of the normalised angular momentum vector, $j_x/j$, of selected stars of model \model{4b} achieving high eccentricity. While the eccentricity evolution seems at first sight compatible with the secular instability in flat eccentric discs as discovered by \citet{mlh08}, the rapid change of the stars' orbital planes as indicated by $j_x/j$ is revealing of Kozai resonance. The former secular instability can only occur in a flat (single, non-warping and non-thickening), eccentric disc in a cusp with a density profile $\rho \propto r^{-1.5}$.

Interestingly, we find the orbital planes eventually flipping at very high eccentricities (e.g.\ solid curve at maximum eccentricity, $t \approx 5.1$\,Myr: $j_x/j$ suddenly changes from $\approx 1$ to $\approx -1$). This was first discovered by \citet{ttk08} who applied their softened version of Gauss' method to the two-disc problem. However, in our case no softening is involved. At such high eccentricities, no large torque is needed to change the small angular momentum to the opposite direction (note that this does not occur in an isolated three-body system undergoing Kozai resonance).

In another calculation (\model{4d}; using the top-heavy IMF with $\alpha = 1.35$), we used the PN terms for all disc stars from the beginning, and a spherical cusp composed of $40\,\msun$ SBHs, which resulted in the creation of 8 B-type and 1 O-type S-star.
Almost all of the originating binaries had close encounters with one of the SBHs (although not close enough to be disrupted),
however none of these encounters lead to a sudden change in orbital eccentricity or semi-major axis which could explain the high eccentricity it achieved much later in the integration. We ascribe the higher number of S-stars created to the fact that the effectiveness of resonant relaxation is proportional to the SBH mass \citep{rt96}. Apparently, pericentre shift due to relativistic effects does not prevent such high eccentricities.

We conclude that if the stellar cusp in the Galactic Centre is mass-dominated by massive SBHs (as predicted by \citealp{fak06} and \citealp{ah08}),
at least some of the S-stars observed could have originated from the young stellar discs.
In a setup of an initially eccentric disc (as suggested by \citealp{lb09} for the counter-clockwise disc), we can expect a somewhat higher probability of high-eccentricity orbits required for the formation of S-stars.
Since the age of the S-stars is less well-determined than that of the stellar discs, some of them may stem from an earlier generation of discs, which would anyway be needed to explain the longer travel times of the hyper-velocity stars observed in the Milky Way halo (\citealp{b07b} and references therein), assuming the same disruption scenario as their origin
\citep{lbk08}.
Furthermore, scattering of young binaries by massive perturbers may contribute to the observed number of S-stars \citep{pha06}.

\section{Conclusions}\label{sec:gcdci:discussion}

In this paper, we have investigated the influence of the stellar cusp on the evolution of young stellar discs in the Galactic Centre.
Our results show that the impact of stars and stellar remnants on the disc stars' orbits must not be neglected in theoretical models and numerical simulations:
The stellar cusp enhances evolution of orbital eccentricities and disc thickness by (non-)resonant relaxation of angular momentum as described
by \citet{lb09}.
We find that this is especially true for high eccentricities, which are prevented by a smooth cusp potential resulting in orbital precession, but are achieved in a grainy cusp as we expect to find it in the inner parsec of our Galaxy. A combination of two-body relaxation processes resulting in a random increase of eccentricity and short-term Kozai resonances leads to eccentricities high enough for binary disruption, leaving behind an S-star on a tight orbit and ejecting a hyper-velocity star. This process is not as efficient as it would be in the absence of a stellar cusp, but a few generations of stellar discs could account for the observed S-stars.

We conclude that the creation of S-stars is a natural consequence of the interaction of two stellar discs in a grainy cusp around an SMBH, where stellar orbits are exposed to the effects of general relativity.
Scattering of stars onto highly eccentric orbits is enhanced by a cusp composed of more massive particles, such as SBH multiples, (remnants of) star clusters or IMBHs. Thus, more realistic modelling of stellar dynamics in the Galactic Centre requires a better understanding of the composition of the stellar cusp.

Even more so, the composition of the stellar discs remains unclear. The mere existence of the S-stars may be regarded as evidence for a steeper (i.e.\ more standard-like) IMF, as this would significantly increase the number of B-stars (and thus stars eventually reaching tight orbits) given the observed number of O-stars. A detailed survey of B-type stars in the central parsec of the Milky Way will not only help us to better understand the composition of the young stellar discs and thus the formation of the S-stars, but also increase the significance of the counter-clockwise disc as well as the observed disc warping.
However, if the formation of stellar discs in the Galactic Centre is a common scenario, most of the B-type stars (including the known S-stars) may stem from previous generations of discs, and their orbits are expected to be randomised by the influence of the cusp as well as the potential of current and previous discs.

As a by-product of our models, we can determine the merger rates of stellar black holes with the SMBH. These mergers can be caused by an SBH plunging into the SMBH, or by gradual inspiral due to gravitational wave emission (so-called extreme mass ratio inspirals, EMRIs), yielding a promising scenario for the detection of gravitational wave signals by the planned Laser Interferometer Space Antenna \citep[e.g.][]{agf+07}.
Since in our models we focus on the evolution of the stellar discs, we do not differentiate between EMRIs and plunges of SBHs, but only identify the merger of an SBH with the SMBH as the point where their distance is smaller than three Schwarzschild radii.
Our models including a cusp of 14\,000 SBHs of mass 15\,\tmsun\ showed merger rates from $20-80$\,Myr$^{-1}$, with a mean of 58\,Myr$^{-1}$.
Note that these rates may be higher when using SBHs beyond 0.22\,pc or of smaller masses (and thus a higher number). On the other hand, they need to be corrected for cusp depletion and loss-cone-refilling.

\section*{Acknowledgments}
We would like to thank Pau Amaro-Seoane for useful comments.
This work was supported by the German Research Foundation (DFG) through the priority program 1177
`Witnesses of Cosmic History: Formation and Evolution of Black Holes, Galaxies and Their Environment'.

\bibliographystyle{aa_mn2e}

\begin{thebibliography}{48}
\expandafter\ifx\csname natexlab\endcsname\relax\def\natexlab#1{#1}\fi

\bibitem[{{Alexander} {et~al.}(2007){Alexander}, {Begelman}, \&
  {Armitage}}]{aba06}
{Alexander} R.~D.,  {Begelman} M.~C.,    {Armitage} P.~J.,  2007, \apj, 654,
  907

\bibitem[{{Alexander}(2005)}]{a05}
{Alexander} T.,  2005, PhR, 419, 65

\bibitem[{{Alexander} \& {Hopman}(2009)}]{ah08}
{Alexander} T.,  {Hopman} C.,  2009, \apj, 697, 1861

\bibitem[{{Amaro-Seoane} {et~al.}(2004){Amaro-Seoane}, {Freitag}, \&
  {Spurzem}}]{afs04}
{Amaro-Seoane} P.,  {Freitag} M.,    {Spurzem} R.,  2004, \mnras, 352, 655

\bibitem[{{Amaro-Seoane} {et~al.}(2007){Amaro-Seoane}, {Gair}, {Freitag},
  {Miller}, {Mandel}, {Cutler}, \& {Babak}}]{agf+07}
{Amaro-Seoane} P.,  {Gair} J.~R.,  {Freitag} M.,  {Miller} M.~C.,  {Mandel} I.,
   {Cutler} C.~J.,    {Babak} S.,  2007, Classical and Quantum Gravity, 24, 113

\bibitem[{{Bahcall} \& {Wolf}(1976)}]{bw76}
{Bahcall} J.~N.,  {Wolf} R.~A.,  1976, \apj, 209, 214

\bibitem[{{Bartko} {et~al.}(2009){Bartko}, {Martins}, {Fritz}, {Genzel},
  {Levin}, {Perets}, {Paumard}, {Nayakshin}, {Gerhard}, {Alexander},
  {Dodds-Eden}, {Eisenhauer}, {Gillessen}, {Mascetti}, {Ott}, {Perrin},
  {Pfuhl}, {Reid}, {Rouan}, {Sternberg}, \& {Trippe}}]{bmf08}
{Bartko} H.,  {et~al.,} 2009, \apj, 697, 1741

\bibitem[{{Baumgardt} {et~al.}(2004{\natexlab{a}}){Baumgardt}, {Makino}, \&
  {Ebisuzaki}}]{bme04a}
{Baumgardt} H.,  {Makino} J.,    {Ebisuzaki} T.,  2004{\natexlab{a}}, \apj,
  613, 1133

\bibitem[{{Baumgardt} {et~al.}(2004{\natexlab{b}}){Baumgardt}, {Makino}, \&
  {Ebisuzaki}}]{bme04b}
{Baumgardt} H.,  {Makino} J.,    {Ebisuzaki} T.,  2004{\natexlab{b}}, \apj,
  613, 1143

\bibitem[{{Blanchet}(2006)}]{bla06}
{Blanchet} L.,  2006, Living Reviews in Relativity, 9, 4

\bibitem[{{Brown} {et~al.}(2007){Brown}, {Geller}, {Kenyon}, {Kurtz}, \&
  {Bromley}}]{b07b}
{Brown} W.~R.,  {Geller} M.~J.,  {Kenyon} S.~J.,  {Kurtz} M.~J.,    {Bromley}
  B.~C.,  2007, \apj, 671, 1708

\bibitem[{{Chang}(2009)}]{chang08}
{Chang} P.,  2009, \mnras, 393, 224

\bibitem[{{Cuadra} {et~al.}(2008){Cuadra}, {Armitage}, \& {Alexander}}]{caa08}
{Cuadra} J.,  {Armitage} P.~J.,    {Alexander} R.~D.,  2008, \mnras, 388, L64

\bibitem[{{Eilon} {et~al.}(2009){Eilon}, {Kupi}, \& {Alexander}}]{eka08}
{Eilon} E.,  {Kupi} G.,    {Alexander} T.,  2009, \apj, 698, 641

\bibitem[{{Freitag} {et~al.}(2006){Freitag}, {Amaro-Seoane}, \&
  {Kalogera}}]{fak06}
{Freitag} M.,  {Amaro-Seoane} P.,    {Kalogera} V.,  2006, \apj, 649, 91

\bibitem[{{Genzel} {et~al.}(1994){Genzel}, {Hollenbach}, \& {Townes}}]{ght94}
{Genzel} R.,  {Hollenbach} D.,    {Townes} C.~H.,  1994, Reports of Progress in
  Physics, 57, 417

\bibitem[{{G\"usten} {et~al.}(1987){G\"usten}, {Genzel}, {Wright}, {Jaffe},
  {Stutzki}, \& {Harris}}]{ggw+87}
{G\"usten} R.,  {Genzel} R.,  {Wright} M.~C.~H.,  {Jaffe} D.~T.,  {Stutzki} J.,
     {Harris} A.~I.,  1987, \apj, 318, 124

\bibitem[{{Hopman} \& {Alexander}(2006)}]{ha06b}
{Hopman} C.,  {Alexander} T.,  2006, \apj, 645, 1152

\bibitem[{{Hurley} {et~al.}(2000){Hurley}, {Pols}, \& {Tout}}]{hpt00}
{Hurley} J.~R.,  {Pols} O.~R.,    {Tout} C.~A.,  2000, \mnras, 315, 543

\bibitem[{{Ivanov} {et~al.}(2005){Ivanov}, {Polnarev}, \& {Saha}}]{ips05}
{Ivanov} P.~B.,  {Polnarev} A.~G.,    {Saha} P.,  2005, \mnras, 358, 1361

\bibitem[{{Kroupa}(2001)}]{k01}
{Kroupa} P.,  2001, \mnras, 322, 231

\bibitem[{{Kroupa}(2008)}]{k08}
{Kroupa} P.,  2008, in Lecture Notes in Physics, Vol. 760, The Cambridge N-Body
  Lectures, ed. S.~J. {Aarseth}, C.~A. {Tout}, \& R.~A. {Mardling} (Berlin:
  Springer Verlag), 181

\bibitem[{{Levin} \& {Beloborodov}(2003)}]{lb03}
{Levin} Y.,  {Beloborodov} A.~M.,  2003, \apjl, 590, L33

\bibitem[{{L{\"o}ck\-mann} \& {Baumgardt}(2008)}]{lb08}
{L{\"o}ck\-mann} U.,  {Baumgardt} H.,  2008, \mnras, 384, 323

\bibitem[{{L{\"o}ckmann} \& {Baumgardt}(2009)}]{lb09}
{L{\"o}ckmann} U.,  {Baumgardt} H.,  2009, \mnras, 394, 1841

\bibitem[{{L{\"o}ckmann} {et~al.}(2008){L{\"o}ckmann}, {Baumgardt}, \&
  {Kroupa}}]{lbk08}
{L{\"o}ckmann} U.,  {Baumgardt} H.,    {Kroupa} P.,  2008, \apjl, 683, L151

\bibitem[{{Lu} {et~al.}(2006){Lu}, {Ghez}, {Hornstein}, {Morris}, {Matthews},
  {Thompson}, \& {Becklin}}]{lu06}
{Lu} J.~R.,  {Ghez} A.~M.,  {Hornstein} S.~D.,  {Morris} M.,  {Matthews} K.,
  {Thompson} D.~J.,    {Becklin} E.~E.,  2006, Journal of Physics Conference
  Series, 54, 279

\bibitem[{{Lu} {et~al.}(2009){Lu}, {Ghez}, {Hornstein}, {Morris}, {Becklin}, \&
  {Matthews}}]{lu09}
{Lu} J.~R.,  {Ghez} A.~M.,  {Hornstein} S.~D.,  {Morris} M.~R.,  {Becklin}
  E.~E.,    {Matthews} K.,  2009, \apj, 690, 1463

\bibitem[{{Madigan} {et~al.}(2009){Madigan}, {Levin}, \& {Hopman}}]{mlh08}
{Madigan} A.-M.,  {Levin} Y.,    {Hopman} C.,  2009, \apjl, 697, L44

\bibitem[{{Maillard} {et~al.}(2004){Maillard}, {Paumard}, {Stolovy}, \&
  {Rigaut}}]{mpsr04}
{Maillard} J.~P.,  {Paumard} T.,  {Stolovy} S.~R.,    {Rigaut} F.,  2004, \aap,
  423, 155

\bibitem[{{Makino} \& {Aarseth}(1992)}]{ma92}
{Makino} J.,  {Aarseth} S.~J.,  1992, PASJ, 44, 141

\bibitem[{{Maness} {et~al.}(2007){Maness}, {Martins}, {Trippe}, {Genzel},
  {Graham}, {Sheehy}, {Salaris}, {Gillessen}, {Alexander}, {Paumard}, {Ott},
  {Abuter}, \& {Eisenhauer}}]{mmt07}
{Maness} H.,  {et~al.,} 2007, \apj, 669, 1024

\bibitem[{{Nayakshin} {et~al.}(2007){Nayakshin}, {Cuadra}, \&
  {Springel}}]{ncs07}
{Nayakshin} S.,  {Cuadra} J.,    {Springel} V.,  2007, \mnras, 379, 21

\bibitem[{{Nayakshin} {et~al.}(2006){Nayakshin}, {Dehnen}, {Cuadra}, \&
  {Genzel}}]{ndcg06}
{Nayakshin} S.,  {Dehnen} W.,  {Cuadra} J.,    {Genzel} R.,  2006, \mnras, 366,
  1410

\bibitem[{{Nayakshin} \& {Sunyaev}(2005)}]{ns05}
{Nayakshin} S.,  {Sunyaev} R.,  2005, \mnras, 364, L23

\bibitem[{{Paumard} {et~al.}(2006){Paumard}, {Genzel}, {Martins}, {Nayakshin},
  {Beloborodov}, {Levin}, {Trippe}, {Eisenhauer}, {Ott}, {Gillessen}, {Abuter},
  {Cuadra}, {Alexander}, \& {Sternberg}}]{pau06}
{Paumard} T.,  {et~al.,} 2006, \apj, 643, 1011

\bibitem[{{Perets} {et~al.}(2008){Perets}, {Gualandris}, {Merritt}, \&
  {Alexander}}]{pgma08}
{Perets} H.~B.,  {Gualandris} A.,  {Merritt} D.,    {Alexander} T.,  2008,
  Memorie della Societa Astronomica Italiana, 79, 1100

\bibitem[{{Perets} {et~al.}(2007){Perets}, {Hopman}, \& {Alexander}}]{pha06}
{Perets} H.~B.,  {Hopman} C.,    {Alexander} T.,  2007, \apj, 656, 709

\bibitem[{{Preto} {et~al.}(2004){Preto}, {Merritt}, \& {Spurzem}}]{pms04}
{Preto} M.,  {Merritt} D.,    {Spurzem} R.,  2004, \apjl, 613, L109

\bibitem[{{Rauch} \& {Tremaine}(1996)}]{rt96}
{Rauch} K.~P.,  {Tremaine} S.,  1996, New Astronomy, 1, 149

\bibitem[{{Salpeter}(1955)}]{s55}
{Salpeter} E.~E.,  1955, \apj, 121, 161

\bibitem[{{Sch{\"o}del} {et~al.}(2007){Sch{\"o}del}, {Eckart}, {Alexander},
  {Merritt}, {Genzel}, {Sternberg}, {Meyer}, {Kul}, {Moultaka}, {Ott}, \&
  {Straubmeier}}]{sea+07}
{Sch{\"o}del} R.,  {et~al.,} 2007, \aap, 469, 125

\bibitem[{{Sch{\"o}del} {et~al.}(2009){Sch{\"o}del}, {Merritt}, \&
  {Eckart}}]{sme09}
{Sch{\"o}del} R.,  {Merritt} D.,    {Eckart} A.,  2009, \aap, accepted
  (astro-ph/0902.3892)

\bibitem[{{Sellwood} \& {Wilkinson}(1993)}]{sw93}
{Sellwood} J.~A.,  {Wilkinson} A.,  1993, Reports on Progress in Physics, 56,
  173

\bibitem[{{Sheth} {et~al.}(2008){Sheth}, {Elmegreen}, {Elmegreen}, {Capak},
  {Abraham}, {Athanassoula}, {Ellis}, {Mobasher}, {Salvato}, {Schinnerer},
  {Scoville}, {Spalsbury}, {Strubbe}, {Carollo}, {Rich}, \& {West}}]{see+08}
{Sheth} K.,  {et~al.,} 2008, \apj, 675, 1141

\bibitem[{{Touma} {et~al.}(2009){Touma}, {Tremaine}, \& {Kazandjian}}]{ttk08}
{Touma} J.~R.,  {Tremaine} S.,    {Kazandjian} M.~V.,  2009, \mnras, 194

\bibitem[{{{\v S}ubr} {et~al.}(2008){{\v S}ubr}, {Schovancov\'a}, \&
  {Kroupa}}]{ssk08}
{{\v S}ubr} L.,  {Schovancov\'a} J.,    {Kroupa} P.,  2008, \aap, accepted
  (astro-ph/0812.1567)

\bibitem[{{Yu} \& {Tremaine}(2003)}]{yt03}
{Yu} Q.,  {Tremaine} S.,  2003, \apj, 599, 1129

\end{thebibliography}

\makeatletter   \renewcommand{\@biblabel}[1]{[#1]}   \makeatother

\label{lastpage}

\end{document}